\documentclass[fleqn,twoside]{article}
\usepackage{espcrc2}
\usepackage{graphicx,epsf}
\usepackage{dcolumn}
\usepackage{bm}

\title{Upper limits on neutrino masses from cosmology}
\author{{\O}ystein Elgar{\o}y\address[ITA]{Institute of theoretical astrophysics, University of Oslo,\\ P.O. Box 1029, N-0315 Oslo, Norway}}

\begin{document}
\begin{abstract}
Upper limits on neutrino masses from cosmology have been reported recently to reach the impressive sub-eV level, which is competitive with future terrestrial neutrino experiments.  In this brief overview of 
the latest limits from cosmology I point out some of the caveats that 
should be borne in mind when interpreting the significance of these limits. 
\end{abstract}

\maketitle

\section{Introduction}

The latest results from the WMAP satellite \cite{spergel} confirm the success 
of the $\Lambda$CDM model, where $\sim 75\;\%$ of the mass-energy density is in the form of dark energy, with matter, most of it in the form of cold dark matter (CDM)  making up the remaining 25 \% .   Neutrinos with masses on the 
eV scale or below will be a hot component of the dark matter and 
will free-stream out of overdensities and thus wipe out small-scale 
structures.  This fact makes it possible to use observations of the clustering 
of matter in the universe to put upper bounds on 
the neutrino masses.  An excellent review of the subject can be found in 
\cite{lesgourgues}. 
With the improved quality of cosmological data 
sets seen in recent years, the upper limits have improved, and some 
quite impressive claims have been made in the recent literature. 
I will in the following summarize the latest upper bounds and point out 
some of the potential systematic uncertainties that need to be clarified 
in the future.

\section{Current constraints on neutrino masses}

One of the assumptions underlying cosmological neutrino mass bounds 
is that neutrinos have no non-standard interactions and that they decouple 
from the thermal background at temperatures of order 1 MeV.  In that 
case, the relation between the sum of the neutrino masses 
$M_\nu$ and their contribution to the energy density of the universe 
is given by 
$\Omega_\nu h^2 = M_\nu / 93.14\;{\rm eV}$
\cite{mangano},
where $h$ is the dimensionless Hubble parameter defined by writing 
the Hubble constant as $H_0 = 100h\;{\rm km}\,{\rm s}^{-1}\,{\rm Mpc}^{-1}$.
The 95\% confidence upper bounds range from $\sim 2\;{\rm eV}$
from the CMB alone (\cite{fukugita}, see also \cite{jostein1}) to $0.17\;{\rm eV}$ 
from a fit to CMB, SNIa, and large-scale structure, including 
the Lyman $\alpha$ forest \cite{seljak}.   
The tightest limits in the table are quite impressive, and with limits 
reaching deep into the sub-eV region it is prudent to point out that 
there are systematic uncertainties associated with these limits.  
I will in the following discuss briefly two 
different types of uncertainties: cosmological uncertainties 
(models and priors) and astrophysical uncertainties 
(e.g. galaxy-dark matter bias).  
\begin{table}[ht]
\begin{center}
\begin{tabular}{l|l}
\hline
Reference & $M_\nu$-bound\\
\hline
Elgar\o y et al.02\cite{elgaroy} &  $1.8$ eV\\
S\'{a}nchez et al. 05\cite{sanchez}&  $1.2$ eV\\
Goobar et al. 06\cite{goobar} & $0.62$ eV \\ 
Fukugita et al. 06\cite{fukugita} &  $2.0$ eV \\
Spergel et al. 06\cite{spergel} & 0.68 eV \\  
Seljak et al. 06\cite{seljak}&  $0.17$ eV \\
Kristiansen et al. 06\cite{jostein2}& $1.4$ eV 
\end{tabular}
\end{center}
\caption{Some recent cosmological neutrino mass bounds (95 \% CL).}\label{boundtable}
\end{table}

\section{Uncertainties in the underlying model}
Most of the limits in table \ref{boundtable} are derived under the 
assumption that the underlying cosmological 
model is the standard spatially flat $\Lambda$CDM model with adiabatic 
primordial perturbations.  Slight variants have been considered, 
e.g. running spectral index of the primordial perturbation spectrum 
\cite{seljak} and varying equation of state parameter $w$ for the 
dark energy component \cite{goobar,steen}.  A significant degeneracy 
between $w$ and $M_\nu$ was pointed out by Hannestad \cite{steen}. 
This degeneracy is a result of the fact that the 
matter power spectrum depends on $f_\nu = \Omega_\nu / \Omega_{\rm m}$, 
and allowing $w$ to vary weakens the constraints on $\Omega_{\rm m}$ and 
hence indirectly on $M_\nu \propto f_\nu \Omega_{\rm m}$. 
The degeneracy can be broken by including constraints from e.g. 
baryon acoustic oscillations \cite{baryons} as shown in \cite{goobar}, 
or, quite simply, by using better data (\cite{jostein1,zunckel}).

The spatially flat 
$\Lambda$CDM model describes the existing cosmological data well, 
but we are not yet in a position to exclude significant variations.  
For example, models where gravity is different from standard 
Einstein gravity are still viable, and that might change the 
neutrino mass limit significantly (see e.g. \cite{david}) for an 
example).  

The CMB is a very clean cosmological probe in the sense that the 
the extraction of the anisotropy signal from the data involves 
relatively few and well justified astrophysical assumptions. 
From this point of view the upper mass bound of 2 eV at the 
95 \% confidence level from the CMB data alone found by Fukugita 
et al. \cite{fukugita} is the most robust one. 
However, if one extends the space of models investigated to 
models that are very different from $\Lambda$CDM, there is no 
upper bound on neutrino masses from the CMB.  To demonstrate this 
point, I point out that Blanchard et al. 
\cite{blanchard} found that an Einstein-de Sitter model with 
$M_\nu=2.4\;{\rm eV}$ gave an excellent fit to the WMAP data, 
provided that there are oscillations in the primordial power 
spectrum, as produced e.g. if there is a phase transition 
during inflation.  Now this model has both a low 
Hubble parameter ($h=0.46$), no cosmological constant (and hence 
a bad fit to the supernovae type Ia data), and is also in some 
tension with the baryonic acoustic oscillations, but this goes 
to show that the CMB alone cannot produce a constraint on the 
neutrino mass once one allows for more radical departures from 
$\Lambda$CDM.

To get really tight constraints on the neutrino masses, one 
needs to include large-scale structure data, since the CMB alone  
cannot go much below 2 eV in sensitivity.  It is, however, 
worth nothing that large-scale structure alone cannot do the 
job.  Take the 2dFGRS power spectrum as an example.  
Figure {\ref{fig:fig1} shows the power spectrum of the 
full 2dFGRS survey as determined by Cole et al \cite{cole}.  Figure 1 also shows the power spectra for a model 
with no massive neutrinos and $\Omega_{\rm m}h=0.168$, 
and for a model with $f_\nu = 0.18$ and $\Omega_{\rm m}h=0.38$.  
These two models have identical values of the $\chi^2$ for the 
data in the range used in fits, marked by the dashed vertical lines 
in the figure.  Thus, one can still hide a lot of neutrinos in 
the matter power spectrum when one does not make use of external 
constraints on $\Omega_{\rm m}$.

\begin{figure}
\includegraphics[width=6cm,height=6cm,angle=-90]{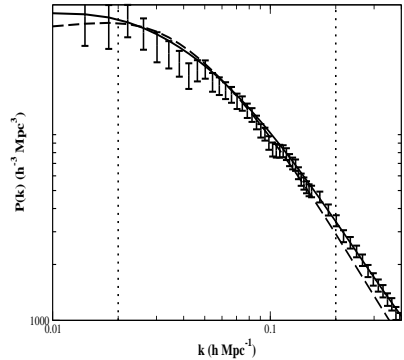}
\caption{\label{fig:fig1}The power spectrum of the full 2dFGRS (bars) 
along with models with $f_\nu=0,\,\Omega_{\rm m}h=0.168$  (full line) 
and $f_\nu=0.18,\,\Omega_{\rm m}h=0.38$ (dashed line). }
\end{figure}

\section{Astrophysical systematics: bias}

Galaxy redshift surveys measure the distribution of galaxies in the 
local universe.  The relation between the distribution of the luminous 
matter and the dark matter is therefore an important issue when 
estimating cosmological parameters in general, and the neutrino mass 
in particular (see e.g. \cite{elglahav} for an overview). 
The usual assumption is that the power spectrum of the matter distribution 
is proportional to the galaxy power spectrum on scales in the linear regime, 
so-called constant bias.  However, recent results 
\cite{sanchez,jostein2,cole2} indicate that there are inconsistencies 
between the results obtained by combining the WMAP data with the 
2dFGRS power spectrum, and those obtained by combining WMAP with 
the SDSS power spectrum obtained in \cite{tegmark}. 
One of the possible interpretations of this 
result is that the bias is scale-dependent \cite{cole2}.  
In any case, this is an important issue to clarify.  
Ideally, one would like to have a direct probe of the mass power spectrum.  
The cluster mass function, with cluster masses derived directly from weak 
gravitational lensing is one such probe, and was obtained for 
the first time in\cite{dahle}.  In \cite{jostein2} this mass 
function was combined with the WMAP 3-year data to obtain a 
95\% upper limit $M_\nu < 1.4\;{\rm eV}$.  Although not as impressive 
as other bounds, this limit is robust in the sense that no assumptions 
about bias were made in deriving it.

\section{Summary}

Current cosmological observations provide strong upper limits on the 
sum of the neutrino masses.  However, when assessing the significance 
of these limits one should bear in mind that several assumptions are 
involved in deriving these limits, both cosmological and astrophysical. 
It is an important task for further research to clarify how sensitive the 
results are to these assumptions, and how well they are justified.

\end{document}